\documentclass[11pt, a4paper]{article}
\usepackage[margin=1in]{geometry}
\usepackage{amsmath}
\usepackage{amssymb} 
\usepackage{graphicx}
\usepackage{hyperref}
\usepackage{booktabs}
\usepackage{caption}
\usepackage{pgfplots}
\pgfplotsset{compat=1.18}
\usepackage{pgfplotstable}
\usepackage{subcaption}
\usepackage{cite}
\usepackage{authblk}
\usepgfplotslibrary{groupplots}

\pgfplotsset{
    every axis/.append style={
        axis line style={black},
        tick style={black},
        line width=0.9pt,
        xlabel style={font=\footnotesize\bfseries, text=black},
        ylabel style={font=\footnotesize\bfseries, text=black},
        title style={font=\footnotesize\bfseries, text=black},
        tick label style={font=\scriptsize\bfseries, text=black},
        legend style={font=\scriptsize\bfseries, text=black}
    },
    mycustomplot/.style={
        width=7cm, height=5.5cm,
        xlabel style={font=\footnotesize\bfseries, text=black},
        ylabel style={font=\footnotesize\bfseries, text=black},
        tick label style={font=\scriptsize\bfseries, text=black},
        title style={font=\footnotesize\bfseries, text=black},
        axis line style={black},
        tick style={black},
        grid=major,
        grid style={dashed, gray!30},
        every axis plot/.append style={line width=1pt},
        legend style={
            font=\scriptsize\bfseries,
            text=black,
            at={(0.5,1.1)},
            anchor=south, 
            legend columns=5, 
            draw=none, 
            fill=none
        },
    },
    unifiedlines/.style={
        cycle multi list={
            {blue, mark=*, mark size=1.2pt},       
            {red, mark=square*, mark size=1.2pt},  
            {green!60!black, mark=triangle*, mark size=1.2pt, dashed}, 
            {orange, mark=diamond*, mark size=1.2pt, dotted}, 
            {violet, mark=pentagon*, mark size=1.2pt, dashdotted} 
        }
    }
}

\title{\textbf{The Co-evolution of Costly Signaling and Cooperation in Social Dilemmas}}
\author[1]{Mahdi Abolhasani}
\author[1]{Saman Moghimi-Araghi}
\author[2,3,4]{Mohammad Salahshour\thanks{\texttt{msalahshour@ab.mpg.de}}}
\affil[1]{\small Department of Physics, Sharif University of Technology, Tehran, Iran}
\date{\today}
\affil[2]{\small Department of Collective Behaviour, Max Planck Institute of Animal Behavior, Konstanz, Germany}
\affil[3]{\small Centre for the Advanced Study of Collective Behaviour, University of Konstanz, Konstanz, Germany}
\affil[4]{\small Department of Biology, University of Konstanz, Konstanz, Germany}

\begin{document}
	
	\maketitle
	
	\begin{abstract}
	Costly cooperation and costly signaling are both difficult to reconcile with simple fitness maximization, yet both are common in biological and social systems. We study a model in which agents emit costly signals and condition their actions on the signals they observe. Across the Prisoner's Dilemma (PD), Snowdrift (SD), and Stag Hunt (SH) games, we ask when this coevolutionary process can sustain cooperation and how it changes across well-mixed populations, spatial lattices, and fluctuating strategic environments. The simulations show that signals are selected less by their raw production costs than by the cooperative responses they currently elicit. In well-mixed populations, the mechanism sustains partial cooperation in PD and SD and drives near-complete cooperation in SH. On lattices, cooperation is strengthened further by local assortment. A reduced mean-field analysis explains why average population feedback is already sufficient in SD and SH, but not in the PD. To account for the PD dynamics, the reduced theory must include transient correlations associated with rare signals, inheritance, or spatial clustering. Our results therefore delineate a class of settings in which costly signals persist because they transiently organize cooperative responses and thereby reshape the effective strategic environment.
\end{abstract}
	
	\section{Introduction}
	
	Cooperation and costly signaling are usually introduced as separate evolutionary puzzles. Cooperation is difficult to explain because a cooperator pays a cost that can be exploited by freeriders, while costly signaling is difficult to explain because the signal itself consumes resources and may increase risk. Both therefore look incompatible with simple fitness maximization \cite{hamilton1964,trivers1971,axelrod1981,nowak2006,lehmann2006,randnowak2013,zahavi1975,grafen1990}. Yet both are widespread in biological and social systems, which suggests that the relevant selective pressures are often more structured than a direct cost argument would imply.
	
Theoretical explanations of costly signaling have followed two main lines. In the standard honest-signaling tradition \cite{zahavi1975,grafen1990,johnstonegrafen1993}, signals are costly precisely because cost helps maintain reliability. Under the handicap principle \cite{zahavi1975,grafen1990}, low-quality individuals cannot cheaply mimic high-quality ones. At the same time, later theoretical work has emphasized that honesty depends more generally on the cost of cheating or on underlying trade-offs, rather than on wasteful equilibrium cost alone \cite{lachmann2001,szamado2011,szamado2023}. In related applications to cooperation, costly prosocial acts can function as honest indicators of quality, commitment, or trustworthiness \cite{gintis2001,roberts1998,barclaywiller2007,fehrler2013,raihani2015}. A different tradition, associated with Dawkins and Krebs \cite{dawkinskrebs1978}, treats signaling less as the transmission of truthful information and more as an arena of strategic conflict in which signalers and receivers coevolve in a manipulation-countermanipulation arms race.

	For the evolution of cooperation, diverse mechanisms have been suggested. Kin selection can promote cooperation by channeling benefits toward relatives \cite{hamilton1964}. Direct reciprocity can reward past cooperation in repeated interactions \cite{trivers1971,axelrod1981}. Reputation-based indirect reciprocity can also stabilize cooperation \cite{nowaksigmund1998,nowak2005indirect}, and a substantial literature has examined the assessment rules \cite{ohtsukiiwasa2004,panchanathanboyd2003} and social norms \cite{salahshour2022moral} required to sustain such systems. Population structure can protect clusters of cooperators from invasion \cite{nowak2006,nowakmay1992,ohtsuki2006}, while multilevel selection can favor cooperation when more cooperative groups outperform less cooperative ones \cite{traulsen2006}. Sanctioning can also sustain cooperation, through prosocial punishment \cite{fehrgaechter2002,salahshour2021punishment} and reward-based incentives \cite{cressman2012}. Signaling, including language \cite{salahshour2020language}, tag-based recognition \cite{riolo2001,jansen2006}, and socially shared or symbolic markers \cite{mcelreath2003,efferson2008,chenli2009} can also promote cooperation. Other mechanisms can act independently as well: voluntary participation and commitment \cite{hauert2002volunteering,songhan2025}, partner choice \cite{noehammerstein1994,barclaywiller2007,fu2008partnerchoice}, freedom of choice between public resources \cite{salahshour2021resources}, heterogeneity in institutions and personalities \cite{salahshour2021institutions,salahshour2021personalities}, and ecological resource flows \cite{salahshour2023ecosystems} can each support cooperation. Recent work further suggests that rational decision-making itself can promote cooperation through perceptual rationality \cite{salahshour2025perceptual} and rational reciprocity \cite{salahshourcouzin2025}.

Recently, it has also been suggested that costly signaling can provide an avenue for resolving conflict in strategic interactions and, in this way, promote cooperation \cite{salahshour2019}. This view builds on, and refines, the evolutionary arms-race perspective of costly signals \cite{dawkinskrebs1978}. Rather than treating signals only as attempts to manipulate receivers, it emphasizes that signals can acquire value because they help organize social responses in a population where behavior is conditioned on observable cues \cite{mcnamara2009trust,baker2012kin}. In that sense, it is closer to literatures in which markers become meaningful through shared expectations and coordinated responses than to models in which signals matter only by revealing hidden type \cite{mcelreath2003,efferson2008,chenli2009,salahshour2020language}. In the model introduced by Salahshour \cite{salahshour2019}, signals have no fixed meaning and need not reveal an underlying quality. Their meaning emerges endogenously through the response conventions attached to them. A costly signal is favored when it elicits cooperation, but this advantage is temporary because successful signals also attract exploitation and eventually lose their effectiveness. The result is a moving coevolutionary process in which costly signals can help create, stabilize, and then erode temporary cooperative niches. This is the mechanism whose scope we revisit here.

	We build on the well-mixed model introduced by Salahshour \cite{salahshour2019}. That study showed that costly signals and partial cooperation can coevolve when agents condition behavior on signals, generating a moving ``signaling war'' in which profitable signals spread until they attract exploitation. The earlier formulation allowed strategies to depend on both the focal individual's own signal and the opponent's signal \cite{salahshour2019}, whereas here we restrict attention to the simpler case in which responses depend only on the opponent's signal. This simplification is empirically motivated by cases in which the strategically relevant cue is carried primarily by the partner being assessed. In humans, for example, trust and cooperation can depend on observed group markers and perceived trustworthiness \cite{chenli2009,fehrler2013}. More generally, both human and animal work have examined cases in which individuals condition their responses on social information about others, including observed cooperative tendencies and cues extracted from prior interactions or conflicts \cite{nowaksigmund1998,mcnamara2009trust,baker2012kin}. Focusing on the opponent-conditioned rule therefore isolates the receiver-side discrimination mechanism that is central to the present argument. However, it is still unclear whether the same mechanism extends across distinct social dilemmas, how spatial structure changes its stability, and what ingredients are needed for a reduced mathematical account of why it works in some games but not in others.
	
Here, we address those gaps. We examine the coevolution of signals and cooperation across the three canonical $2\times 2$ social dilemmas: the Prisoner's Dilemma (PD), which belongs to the dominance class; the Snowdrift game (SD), which belongs to the anti-coordination class; and the Stag Hunt (SH), which belongs to the coordination class \cite{platkowski2017,szabofath2007,salahshourcouzin2025}. Together, these games span qualitatively distinct strategic settings within symmetric two-player, two-strategy interactions and therefore provide a compact test of whether the signaling mechanism depends on a specific payoff structure or carries across strategically different contexts. We then extend the analysis in two directions. First, we study a spatial lattice, where local clustering can interact with signal-based partner discrimination. Second, we consider fluctuating strategic environments in which the operative game changes probabilistically across generations. This extension is motivated by the fact that, in realistic settings, social interactions often unfold across multiple institutional or ecological contexts rather than within a single fixed game \cite{salahshour2022moral,salahshour2021resources}. Across these settings, one empirical pattern keeps returning. Signals are not sorted by production cost alone; they rise and fall according to the cooperative responses they elicit.
	
	The results show that the same signal-response process can sustain partial cooperation in PD and SD, but near-complete cooperation in SH, and that spatial structure strengthens cooperation further. The reduced analysis also clarifies an important asymmetry across games. An independence approximation already explains why cooperation can persist in SD and SH, but it cannot sustain cooperation in the PD. To recover the PD dynamics seen in simulations, the mean-field description must retain the higher-order correlations generated by rare signals, inheritance, and spatial assortment.
	
	We therefore interpret signaling here less as a vehicle for honest quality revelation \cite{zahavi1975,grafen1990,gintis2001} and more as a strategic variable whose meaning is created endogenously \cite{riolo2001,jansen2006,salahshour2019}. A signal matters because of the response convention currently attached to it, and that convention is itself under selection \cite{salahshour2019}. The population continually creates, exploits, and erodes temporary cooperative niches \cite{salahshour2019}. This interpretation does not displace honesty-based accounts \cite{zahavi1975,grafen1990,gintis2001} or arms-race accounts \cite{dawkinskrebs1978}; rather, it identifies a complementary regime in which costly signals matter because they temporarily coordinate who receives cooperation, and because the strategic value of that coordination is itself under selection \cite{salahshour2019}.
	
	\section{Materials and methods}
	
	We consider a population of $N$ agents who repeatedly interact in a one-shot two-player game. Each agent $\alpha$ carries two inherited components. The first is a signal-production distribution
	\[
	\mathbf{p}_{\alpha}=(p_{\alpha 1},\dots,p_{\alpha n}), \qquad \sum_{i=1}^{n} p_{\alpha i}=1,
	\]
	where $p_{\alpha i}$ is the probability that agent $\alpha$ emits signal $\sigma_i$. Each signal $\sigma_i$ has a fixed production cost $c_i \in [0,c_{\max}]$, drawn once at the start of the simulation and then kept fixed for all agents and all generations. The second component is a conditional response rule
	\[
	s_{\alpha}:\{\sigma_1,\dots,\sigma_n\}\rightarrow \{C,D\},
	\]
	so that $s_{\alpha}(\sigma_i)=C$ means that agent $\alpha$ cooperates after observing signal $\sigma_i$ from its opponent. In the present model, actions depend on the opponent's signal only.
	This is simpler than the formulation in \cite{salahshour2019}, where behavior depended on both the focal agent's own signal and the opponent's signal.
	
	\subsection{Interactions and Payoffs}
	Agents interact and accumulate payoffs according to one of two population structures.
	
	In a match between agents $\alpha$ and $\gamma$, each player emits a signal sampled from its own distribution, observes the opponent's signal, and then chooses an action according to its response rule. The focal agent's payoff is the game payoff associated with the action pair, minus the cost of the focal agent's own emitted signal. The total payoff $w_{\alpha}$ is the sum of net payoffs over all interactions in that generation.
	
	The interaction structure is one of the following:
	\begin{itemize}
		\item \textbf{Well-Mixed Population:} In each generation, $N$ agents are randomly paired to play a single-shot, two-player game. Each agent participates in exactly one interaction per generation.
		\item \textbf{Structured Population (Lattice).} Agents occupy the vertices of a two-dimensional $L \times L$ square lattice ($N=L^2$) with periodic boundary conditions. In each generation, every agent plays a game with its four immediate von Neumann neighbors.
	\end{itemize}

    In two-player social dilemmas, the four standard payoffs are: $R$ (reward for mutual cooperation), $S$ (sucker's payoff received by a cooperator when the opponent defects), $T$ (temptation payoff to a defector against a cooperator), and $P$ (punishment for mutual defection). The ordering of these values defines the type of game (e.g., Prisoner's Dilemma, Snowdrift, Stag Hunt). The payoff values used here are summarized in Table \ref{tab:payoffs}.
    
\begin{table}[h!]
    \centering
    \caption{Payoff matrices $(R,S,T,P)$ for different social dilemmas. Payoffs are for the row player.}
    \label{tab:payoffs}
    \begin{tabular}{@{}l|l|cccc@{}}
        \toprule
        \textbf{Game} 
        & \textbf{Defining Condition}
        & \textbf{$R$} 
        & \textbf{$S$} 
        & \textbf{$T$} 
        & \textbf{$P$} \\
        \midrule

        \textbf{Prisoner's Dilemma (PD)} 
        & $T > R > P > S$
        & $3$ & $0$ & $5$ & $1$ \\

        \textbf{Snowdrift (SD) Game} 
        & $T > R > S > P$ 
        & $3$ & $1$ & $5$ & $0$ \\

        \textbf{Stag Hunt (SH) Game} 
        & $R > T \ge P > S$
        & $5$ & $0$ & $3$ & $1$ \\
        \bottomrule
    \end{tabular}
\end{table}

	\subsection{Selection and Mutation}
	The population evolves through Wright-Fisher replacement. After payoffs are collected, the entire population is replaced by a new generation of $N$ offspring.
	\begin{itemize}
		\item \textbf{Reproduction:} Each site in the new generation is filled by an offspring from a parent selected from the previous generation. In the well-mixed model, any individual in the population can be chosen as a parent. The probability that agent $\alpha$ is selected is proportional to its fitness,
\[
f_\alpha = \exp(\beta w_\alpha),
\]
where $w_\alpha$ is its total payoff and $\beta$ is the intensity of selection. The corresponding selection probability is
\[
\Pi_\alpha = \frac{\exp(\beta w_\alpha)}{\sum_{\gamma=1}^N \exp(\beta w_\gamma)}.
\]

In the lattice model, reproduction is local: the parent of a site is chosen from the focal agent and its four nearest neighbors, with the same exponential fitness weighting but normalized over that local neighborhood.
		\item \textbf{Inheritance and Mutation:} An offspring inherits the complete genome $(\mathbf{p}_{parent}, s_{parent})$ of its chosen parent, after which mutations may affect signals and strategies independently.
		\begin{itemize}
			\item \textbf{Signal Mutation:} With probability $\nu_{\sigma}$, a random signal index $i$ is chosen and the offspring's signal-production distribution is shifted toward $\sigma_i$ by adding $d$ to that component and then renormalizing. This implements gradual change in signaling propensity.
			\item \textbf{Strategy Mutation:} With probability $\nu_s$, the offspring's response rule is mutated at $n_s$ randomly chosen entries, each of which is reset to either $C$ or $D$ with equal probability. Strategy mutation therefore explores the space of conditional responses more abruptly than signal mutation explores the signal distribution.
		\end{itemize}
	\end{itemize}
	The default parameter values used throughout the paper are listed in Table \ref{tab:params}.
	
	\begin{table}[h!]
		\centering
		\caption{Default simulation parameters.}
		\label{tab:params}
		\begin{tabular}{@{}lll@{}}
			\toprule
			\textbf{Parameter} & \textbf{Symbol} & \textbf{Default Value} \\ \midrule
			Population Size & $N$ & 1800 (Well-Mixed), 1600 (Lattice) \\
			Number of Signals & $n$ & 100 \\
			Intensity of Selection & $\beta$ & 1.0 \\
			Signal Mutation Rate & $\nu_{\sigma}$ & 0.001 \\
			Signal Mutation Strength & $d$ & 0.2 \\
			Strategy Mutation Rate & $\nu_s$ & 0.001 \\
			Number of Strategy Entries Mutated & $n_s$ & 10 \\
			Maximum Signal Cost & $c_{\max}$ & 0.5 \\ \bottomrule
		\end{tabular}
	\end{table}
	
	\section{Results}
	
	We organize the results around four questions: how the mechanism behaves in the three canonical games, how it changes in a spatial population, how it responds to fluctuating strategic environments, and what determines which signals remain prevalent.
	
	\subsection{Canonical Games in Well-Mixed Populations}
	
	\subsubsection{Evolution of Cooperation}
\begin{figure}[htbp]
\centering
\includegraphics[width=\textwidth]{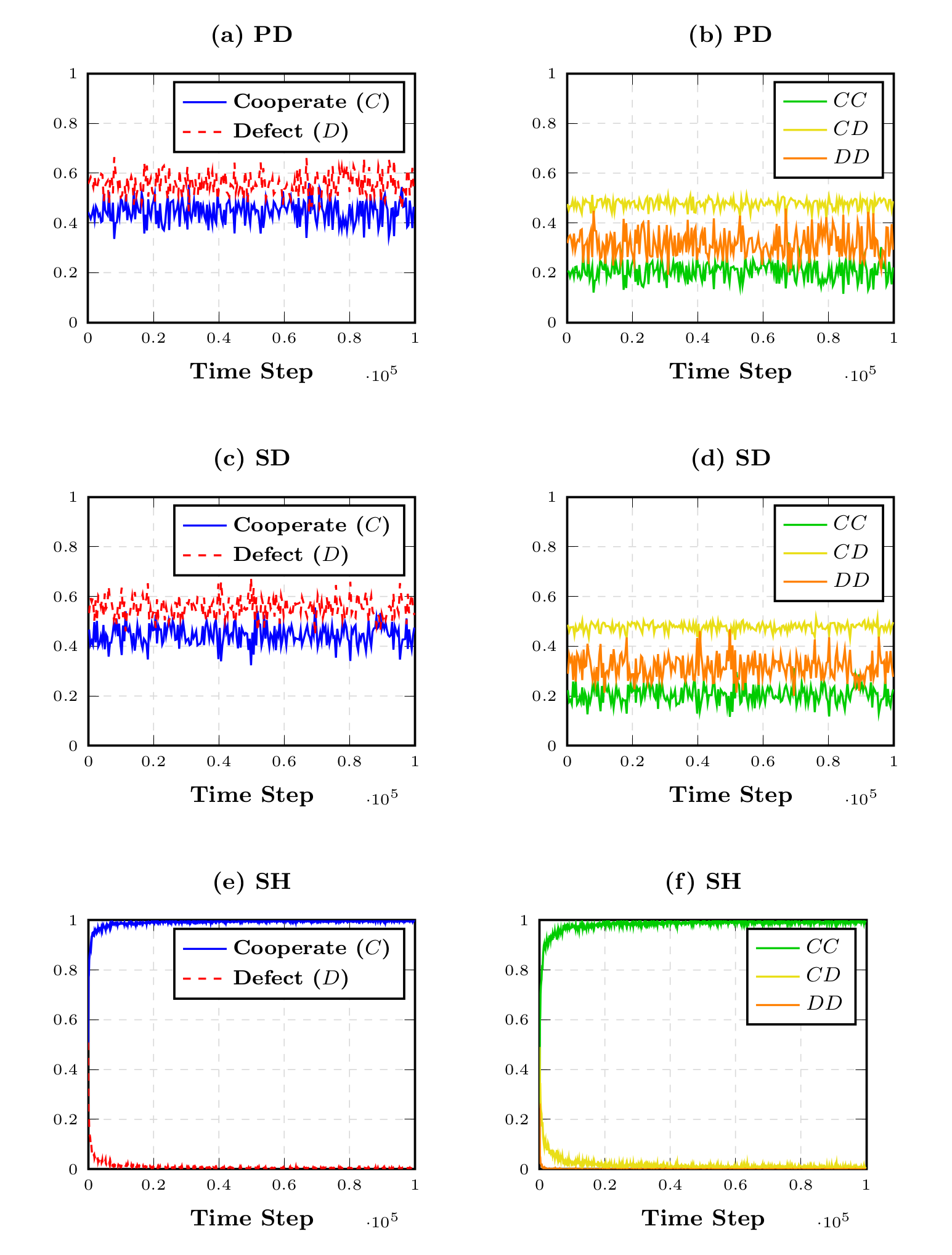}
	\caption{
		Well-mixed action dynamics across the three canonical games. Panels (a,b) show the Prisoner's Dilemma, panels (c,d) the Snowdrift game, and panels (e,f) the Stag Hunt. Panels (a,c,e) report the fraction of cooperative (blue) and defective (red) actions over time, averaged over every 400 generations. Panels (b,d,f) show the frequencies of $CC$, $CD$, and $DD$ outcomes.
		 Parameters: $N=1800$, $n=100$, $\beta=1$, $\nu_{\sigma}=\nu_s=10^{-3}$, $d=0.2$, $n_s=10$, $c_{\max}=0.5$, $10^5$ generations, and game-specific payoff matrices as in Table~\ref{tab:payoffs}.
	}
	\label{fig:wellmixed_actions}
\end{figure}

	We begin with well-mixed populations. In the PD and SD, cooperation survives but does not take over. In Fig.~\ref{fig:wellmixed_actions}(a) and Fig.~\ref{fig:wellmixed_actions}(c), the fraction of cooperative actions remains well above zero, yet defecting actions are still more common. The outcome frequencies in Fig.~\ref{fig:wellmixed_actions}(b) and Fig.~\ref{fig:wellmixed_actions}(d) show why: the system spends most of its time in asymmetric $CD$ matches, with $DD$ still common and $CC$ comparatively rare. In both games, then, signaling does not eliminate conflict. Instead, it stabilizes a heterogeneous regime in which cooperation persists, but remains continually exposed to exploitation.
	
	The SH behaves differently. In Fig.~\ref{fig:wellmixed_actions}(e), cooperation rises quickly toward fixation, and Fig.~\ref{fig:wellmixed_actions}(f) shows the corresponding dominance of $CC$. The contrast with PD and SD is mechanistically informative. Once SH is pushed across its coordination threshold, signaling no longer has to hold cooperation against persistent exploitation. Instead, it mainly helps the population settle on the cooperative equilibrium. These results show that the same signal-response mechanism leads to different social outcomes depending on the underlying game: partial cooperation in PD and SD, but near-complete coordination in SH.

In the Supplementary Information, Fig.~S1, we show that this strategic pattern is robust but not rigid. Across most parameter changes, the well-mixed PD remains in a partially cooperative regime with substantial weight on asymmetric $CD$ interactions. The clearest departures occur when the reward for mutual cooperation is increased, which pushes the system toward $CC$, and when selection becomes too strong, which makes exploitative responses more effective and raises $DD$. Lower strategy mutation and moderate signal mutation are especially favorable because they preserve useful response rules while still allowing the population to discover profitable new signals. See Supplementary Information, Section~S1, for more detail.

\subsubsection{Evolution of Costly Signals}

\begin{figure}[htbp]
\centering
\includegraphics[width=0.96\textwidth]{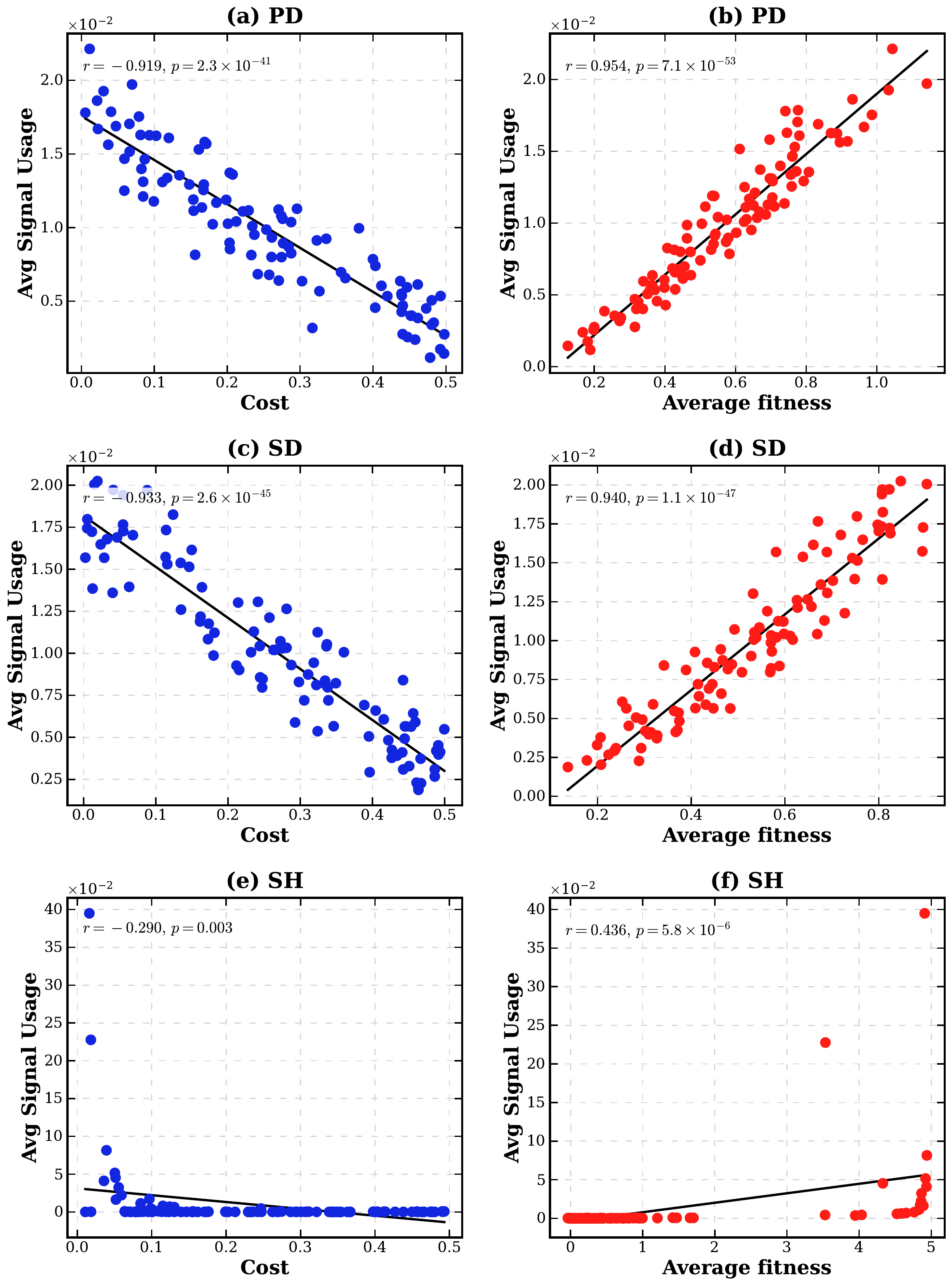}
	\caption{
		Well-mixed signal statistics across the three canonical games. Panels (a,b) show the Prisoner's Dilemma, panels (c,d) the Snowdrift game, and panels (e,f) the Stag Hunt. Panels (a,c,e) plot average signal usage against signal cost. Panels (b,d,f) plot average signal usage against average signal fitness, defined as benefit minus cost. The black line in each panel is the least-squares trend, and the inset reports the Pearson correlation coefficient and its two-sided $p$ value.
		 Parameters: $N=1800$, $n=100$, $\beta=1$, $\nu_{\sigma}=\nu_s=10^{-3}$, $d=0.2$, $n_s=10$, $c_{\max}=0.5$, $10^5$ generations, and game-specific payoff matrices as in Table~\ref{tab:payoffs}.
	}
	\label{fig:wellmixed_signals}
\end{figure}

When we first look at signal frequency as a function of cost, the familiar puzzle of costly signaling is immediately visible. Costly signals do evolve. In Fig.~\ref{fig:wellmixed_signals}(a), Fig.~\ref{fig:wellmixed_signals}(c), and Fig.~\ref{fig:wellmixed_signals}(e), signals with nonzero costs still reach appreciable frequencies, so the dynamics cannot be reduced to simple minimization of signaling cost. A closer look at signal fitness, defined here as the average net payoff associated with emitting a signal, reveals the organizing principle more clearly. In Fig.~\ref{fig:wellmixed_signals}(b), Fig.~\ref{fig:wellmixed_signals}(d), and Fig.~\ref{fig:wellmixed_signals}(f), more frequent signals tend to be those with higher average fitness. This already resolves much of the apparent puzzle. Costly signals do not persist because selection ignores cost, but because some signals generate enough cooperative return to offset that cost and remain favorable in net terms.

The differences across games are also informative. In PD and SD, strategic conflict remains relatively strong. Signals that attract cooperation also create continuing opportunities for exploitation, which is reflected in the persistent weight of asymmetric $CD$ outcomes in Fig.~\ref{fig:wellmixed_actions}(b) and Fig.~\ref{fig:wellmixed_actions}(d). No signal-response convention remains secure for long. As a result, selection repeatedly shifts toward newly favorable signals, and both signal fitness and signal frequencies remain comparatively dispersed in Fig.~\ref{fig:wellmixed_signals}(a)--(d). The resulting turnover is visible in Supplementary Information, Fig.~S4, where the dominant signal bands continue to move over time. 

In SH, by contrast, the conflict is weaker. Once a cooperative convention becomes common, mutual cooperation is more attractive than exploiting it, as seen in the dominance of $CC$ in Fig.~\ref{fig:wellmixed_actions}(e) and Fig.~\ref{fig:wellmixed_actions}(f). The incentive to escape exploitation by shifting toward new, weakly established signals therefore does not exist. Instead, signals mainly serve to coordinate behavior on an already favorable equilibrium. Consequently, signal frequency becomes concentrated and tends to settle on a narrower, cheaper, and fitter part of signal space in Fig.~\ref{fig:wellmixed_signals}(e) and Fig.~\ref{fig:wellmixed_signals}(f).

Cost therefore matters, but not in isolation. Its evolutionary effect is mediated by the response convention attached to the signal and by the strategic conflict surrounding that convention. Consequently, signal fitness, defined as the net benefit of emitting a signal (benefit minus cost), is often a better predictor of signal frequency than its apparent cost.

In the Supplementary Information, Figs.~S2 and S3, we show that this conclusion survives parameter variation. The strongest regularity is that signal prevalence follows signal fitness more reliably than it follows signal cost. As $\nu_{\sigma}$ rises from very small values, the dependence of prevalence on cost first sharpens because the population explores signal space more effectively. At very high $\nu_{\sigma}$, however, the cost relation weakens again even though the fitness relation remains strong. The reason is that signals then turn over faster than strategies, so the response landscape changes only slowly relative to signal frequencies. Selection therefore sorts signals mainly by their current net payoff, while rapid mutation continually reintroduces both cheap and costly variants and blurs any simple dependence on cost alone. At the opposite extreme of very slow signal mutation, exploration is too weak and incumbency matters, so cost again becomes a poor predictor of prevalence. See Supplementary Information, Section~S2, for details.

\subsection{Canonical Games in Structured Populations}

	\subsubsection{Evolution of Cooperation}
\begin{figure}[htbp]
\centering
\includegraphics[width=0.95\textwidth]{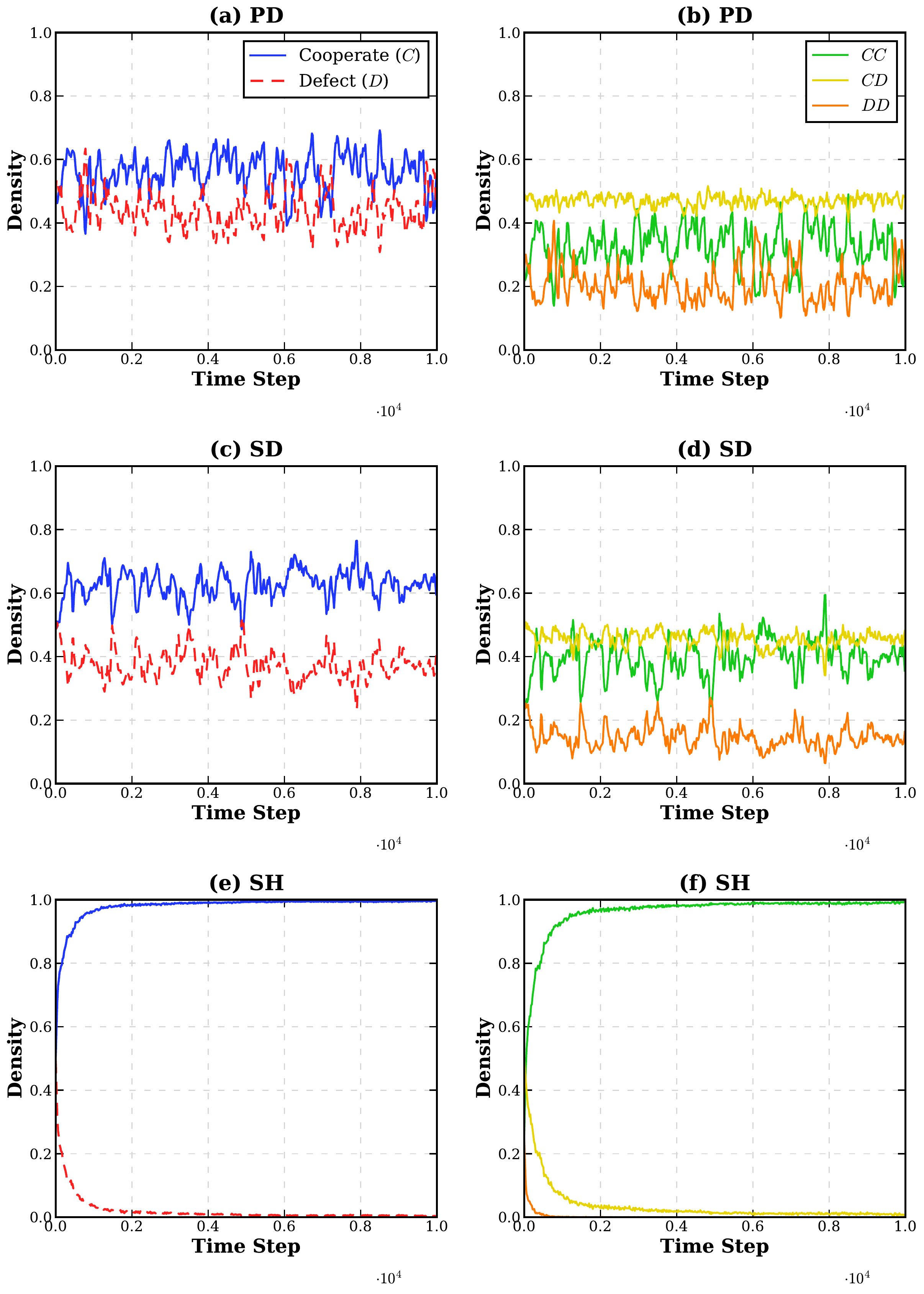}
	\caption{
		Structured-population action and outcome statistics across the three canonical games. Panels (a,b) show the Prisoner's Dilemma, panels (c,d) the Snowdrift game, and panels (e,f) the Stag Hunt. In each pair, the first panel plots the frequencies of cooperative and defective actions over time, and the second panel shows the corresponding outcome frequencies $CC$, $CD$, and $DD$.
		 Parameters: $L=40$ ($N=1600$), $n=100$, $\beta=1$, $\nu_{\sigma}=\nu_s=10^{-3}$, $d=0.2$, $n_s=10$, $c_{\max}=0.5$, $10^4$ generations, and game-specific payoff matrices as in Table~\ref{tab:payoffs}.
	}
	\label{fig:lattice_actions}
\end{figure}

	To examine how signaling interacts with network reciprocity, we repeated all three canonical games on a $40\times 40$ lattice (see Supplementary Information, Fig.~S6, for parameter dependence). The results are summarized in Fig.~\ref{fig:lattice_actions} and Fig.~\ref{fig:lattice_signals}. In PD and SD, cooperation rises substantially. This can be seen in Fig.~\ref{fig:lattice_actions}(a)--(d), where $CC$ becomes more common and $DD$ is reduced. In SH, as in the well-mixed population, cooperation already becomes dominant, so spatial structure mainly stabilizes rather than transforms the outcome.
	
These results show that network reciprocity and strategic signaling reinforce one another, but the mechanism is not simply additive. In the classical snowdrift game without signaling, spatial structure often inhibits rather than promotes cooperation \cite{hauertdoebeli2004}. The opposite pattern here shows that once signaling is coupled to action, space changes more than encounter frequencies. It helps local signal-response conventions form, persist, and resist immediate invasion. Within cooperative neighborhoods, agents repeatedly meet others who attach similar meanings to locally common signals, which raises the payoff to conditional cooperation. At cluster boundaries, exploitative or mismatched neighbors are less likely to trigger the same response. Population structure can therefore promote cooperation in SD not by clustering unconditional cooperators, but by stabilizing local coordination over who is treated as a cooperative partner.

\subsubsection{Evolution of Costly Signals}
\begin{figure}[htbp]
\centering
\includegraphics[width=0.96\textwidth]{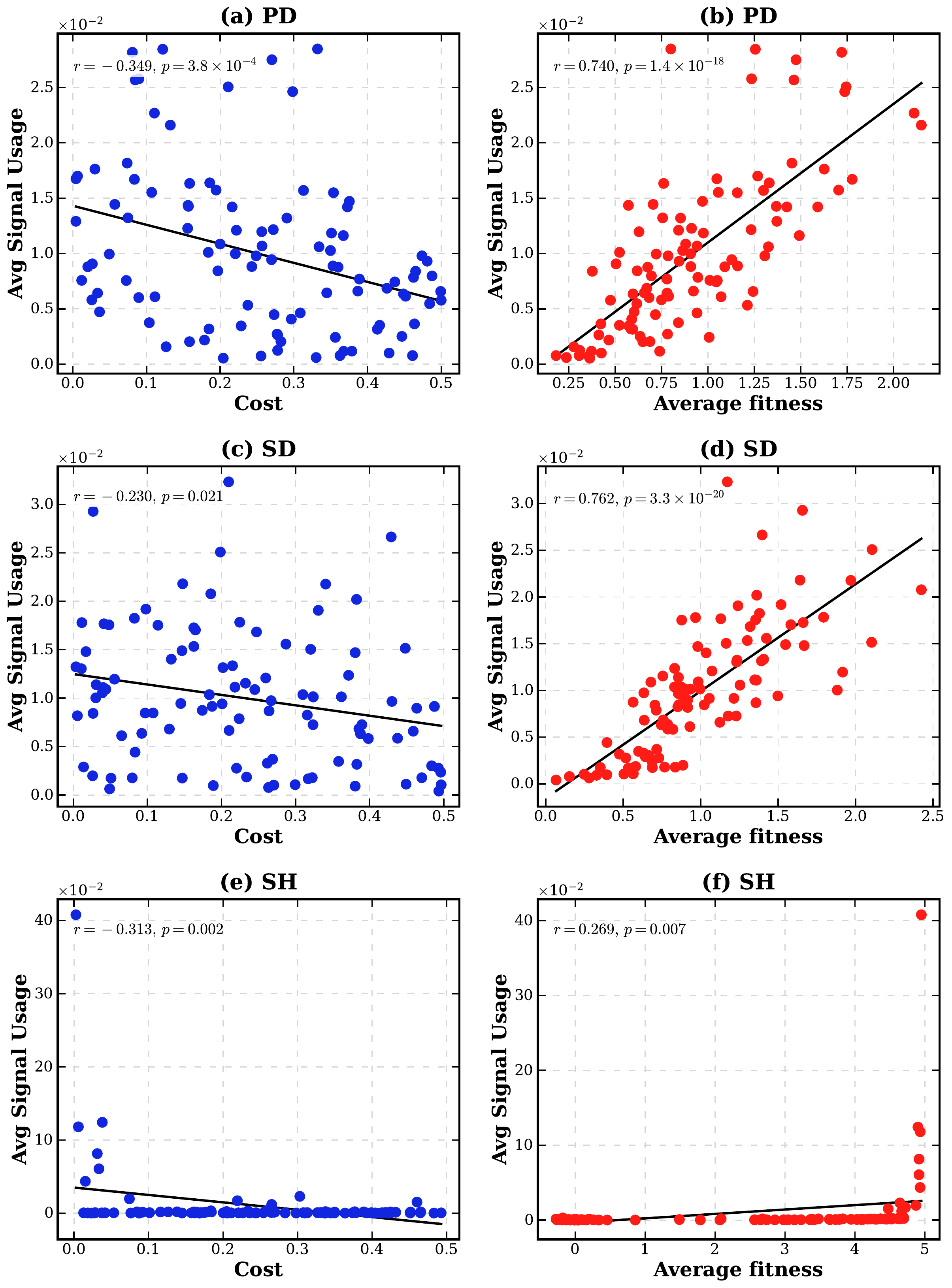}
	\caption{
		Structured-population signal statistics across the three canonical games. Panels (a,b) show the Prisoner's Dilemma, panels (c,d) the Snowdrift game, and panels (e,f) the Stag Hunt. Panels (a,c,e) plot average signal usage against signal cost. Panels (b,d,f) plot average signal usage against average signal fitness, defined as benefit minus cost. The black line in each panel is the least-squares trend, and the inset reports the Pearson correlation coefficient and its two-sided $p$ value.
		 Parameters: $L=40$ ($N=1600$), $n=100$, $\beta=1$, $\nu_{\sigma}=\nu_s=10^{-3}$, $d=0.2$, $n_s=10$, $c_{\max}=0.5$, $10^4$ generations, and game-specific payoff matrices as in Table~\ref{tab:payoffs}.
	}
	\label{fig:lattice_signals}
\end{figure}

In Fig.~\ref{fig:lattice_signals} we present signal frequencies as a function of cost and fitness in a structured population across games. As in the well-mixed population, more frequent signals tend to be cheaper and to have higher average fitness, as seen by comparing Fig.~\ref{fig:lattice_signals} with Fig.~\ref{fig:wellmixed_signals}. But the lattice clouds are noticeably more scattered, especially in PD and SD. This follows naturally from the fact that in structured populations costly signaling is no longer the only protection for cooperation. Network reciprocity already supplies a baseline shelter, so a signal's success depends not only on its population-wide strategic value but also on which neighborhood carries it and which local response rules happen to co-occur with it. Signals can therefore persist by hitchhiking with successful local lineages even when they are not globally strongest, which broadens the structured distributions. In Supplementary Information, Fig.~S5, the structured PD and SD heatmaps show patchier but longer-lived signal bands than in the well-mixed case, whereas SH settles rapidly onto a narrow low-cost band. This indicates that local neighborhoods prolong competition among alternative signals in PD and SD, but mainly reinforce a single cooperative convention in SH once coordination is established.

In the Supplementary Information, Fig.~S6, we explore the effect of parameter values and show that higher cooperation than in the well-mixed case is observed across most of the explored space.

\subsection{\texorpdfstring{Adaptation in Fluctuating Strategic Environments}{Adaptation in Fluctuating Strategic Environments}}

\begin{figure}[htbp]
\centering
\includegraphics[width=\textwidth]{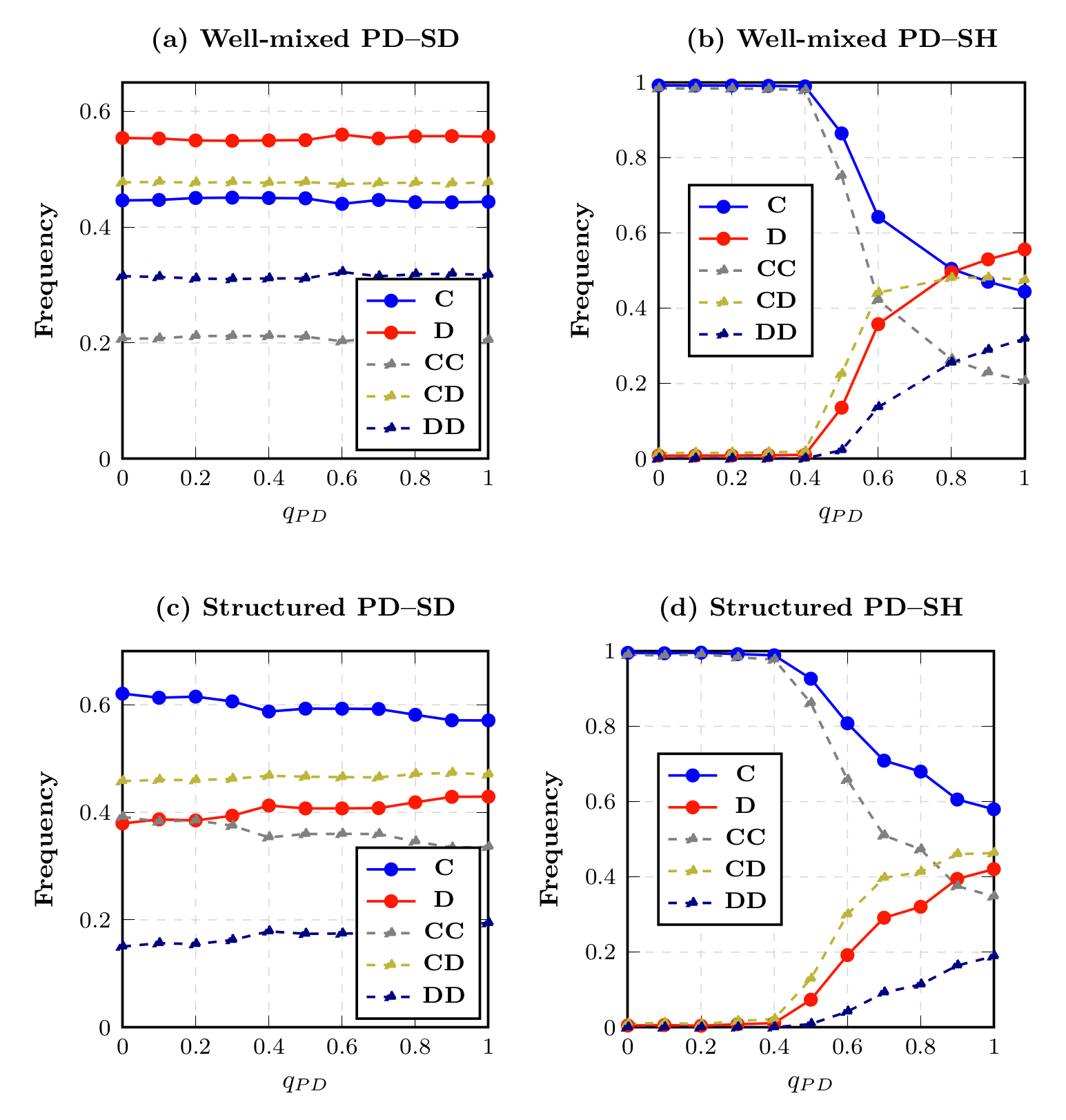}
	\caption{
		Adaptation in fluctuating strategic environments across population structures. The equilibrium fractions of cooperation (C), defection (D), and the three outcome classes ($CC$, $CD$, and $DD$) are plotted as functions of the probability of playing a Prisoner's Dilemma ($q_{PD}$). Panels (a,b) show the well-mixed PD--SD and PD--SH transitions, while panels (c,d) show the corresponding structured-population transitions.
		 Parameters: well-mixed runs use $N=1800$ and structured runs use $L=40$ ($N=1600$); otherwise $n=100$, $\beta=1$, $\nu_{\sigma}=\nu_s=10^{-3}$, $d=0.2$, $n_s=10$, $c_{\max}=0.5$, and the component payoff matrices are those in Table~\ref{tab:payoffs}.
	}
	\label{fig:mixed_games}
\end{figure}

So far, we have focused on fixed strategic environments. In many biological and social settings, however, the strategic problem itself can vary over time \cite{salahshour2022moral,salahshour2021resources}. To investigate that possibility, we let the operative game change probabilistically in each encounter and compare both well-mixed and structured populations in Fig.~\ref{fig:mixed_games}. The top row shows well-mixed populations and the bottom row shows the lattice. In these simulations, the game for each interacting pair is drawn randomly between two specified games with the prescribed probability of playing PD. 

In a well-mixed population, when the fluctuating strategic environment transitions between PD and SD, cooperation remains stable and shows little variation as the probability of playing PD increases. This suggests that the same qualitative conflict-resolution process operates in both games: individuals remain in a signaling arms race, but the signals that spread are those that elicit more favorable strategic responses. Changing the mixture therefore alters the quantitative balance of payoffs more than the logic of the signal-response regime itself. 

In a structured population, cooperation is higher overall. As in the well-mixed case, it remains largely stable as the probability of playing PD increases. Here, however, cooperation declines slightly as the probability of PD rises, indicating that population structure remains more favorable for the evolution of cooperation in SD than in PD. Even so, the absence of a sharp transition shows that the local conventions sustaining cooperation in the two games are compatible enough to carry over across the fluctuating environment.

The combination of PD and SH produces a different pattern. In both the well-mixed and structured populations shown in Fig.~\ref{fig:mixed_games}(b) and Fig.~\ref{fig:mixed_games}(d), cooperation is close to one when SH is common and declines as PD becomes more frequent. The decline is gradual rather than abrupt, and the lattice keeps cooperation appreciably higher deeper into the PD region. The contrast with the PD-SD mixture is informative. PD and SD support broadly the same partially cooperative regime, whereas SH favors near-complete coordination. A PD-SH mixture therefore does not merely vary the strength of a single mechanism; it moves the population between two qualitatively different uses of signaling. In SH, signaling mainly helps select and stabilize a cooperative convention that is already payoff-favored. In PD, the same signals do more demanding work by shielding cooperation from exploitation through conditional discrimination.

The results for fluctuating strategic environments therefore suggest that signal-mediated cooperation is adaptable rather than finely tuned, and that this adaptability survives the addition of spatial structure. When the component games favor similar social states, as in PD-SD, the aggregate outcome is stable because the same partially cooperative signal-mediated regime is effective in both. When they pull in different directions, as in PD-SH, the population shifts continuously between convention selection and dilemma mitigation. The absence of a sharp jump in either topology shows that the same evolving signal system can move smoothly across changing strategic conditions rather than requiring a separate signaling regime for each game.

\subsection{Mean-Field Approach}\label{sec:model_description}
To provide a mathematical understanding of the coevolution of costly signaling and cooperation, we develop two coarse-grained mean-field descriptions. For each signal $\sigma_i$, let $\rho_i(t)$ denote its population frequency at generation $t$, and let $x_i(t)$ denote the fraction of the population that cooperates when facing that signal. The variables $\rho_i$ and $x_i$ therefore summarize, respectively, which signals are common and what response convention the population currently attaches to them. The mean-field analysis is not intended as a faithful reduction of the full agent-based process. Instead, it provides a phenomenological description that separates effects already implied by average population frequencies from those that require correlations between rarity, inheritance, and local assortment. For clarity, we refer to the two reduced descriptions below as the independence closure and the rare-signal-protection closure.

\subsubsection{Independence Closure}\label{sec:independence_closure}
We begin with the simplest closure, in which signal production and response rules are treated as unlinked at the population level. This approximation retains the marginal signal frequencies $\rho_i$ and the response probabilities $x_i$, but neglects higher-order associations between who emits a signal and which response convention is attached to it. It therefore provides a natural baseline: if this closure already reproduces a qualitative result, then that result does not rely on those omitted correlations; if it fails, the missing correlations are part of the explanation.

To reflect the nonoverlapping-generation structure of the simulations, we write the reduced dynamics in discrete time. In the absence of the stochastic terms introduced below, the model reduces to deterministic discrete-time replicator-mutator dynamics \cite{nowak2006}. At generation $t$, the expected payoffs from cooperating and defecting in response to a signal are
\begin{align}
\pi_C(t) &= \sum_{j=1}^{n}\rho_j(t)\big[x_j(t)R+\big(1-x_j(t)\big)S\big], \\
\pi_D(t) &= \sum_{j=1}^{n}\rho_j(t)\big[x_j(t)T+\big(1-x_j(t)\big)P\big].
\end{align}
Using Fermi fitnesses $f_C(t)=e^{\beta\pi_C(t)}$ and $f_D(t)=e^{\beta\pi_D(t)}$, the response profile first evolves by selection,
\[
x_i^{\mathrm{sel}}(t+1)=\frac{x_i(t)f_C(t)}{x_i(t)f_C(t)+\big(1-x_i(t)\big)f_D(t)}.
\]
We then include strategy mutation in the form
\[
x_i^{\mathrm{mut}}(t+1)=\big(1-\mu_s\big)x_i^{\mathrm{sel}}(t+1)+\frac{\mu_s}{2},
\qquad
\mu_s=\nu_s\frac{n_s}{n},
\]
which reflects random reassignment of the conditioned action with probability $\mu_s$. Finally, population stochasticity is added as
\begin{equation}
x_i(t+1)=x_i^{\mathrm{mut}}(t+1)+\sigma_x\sqrt{\frac{x_i^{\mathrm{mut}}(t+1)\big(1-x_i^{\mathrm{mut}}(t+1)\big)}{N}}\,\xi_{x,i}(t),
\label{eq:mf_x_base}
\end{equation}
where $\xi_{x,i}(t)$ is a standard normal draw and $\sigma_x$ controls the strength of finite-population fluctuations. In the numerical implementation, values are truncated back to the interval $[0,1]$ after this stochastic step.
This form is used because, in a finite population with nonoverlapping generations, random sampling around the expected post-mutation frequency has leading-order variance proportional to $x_i^{\mathrm{mut}}(t+1)\big(1-x_i^{\mathrm{mut}}(t+1)\big)/N$. The stochastic term therefore provides a Gaussian diffusion approximation to Wright--Fisher-type demographic noise rather than introducing an additional strategic effect \cite{imhof2006}.

Given the updated response profile, the expected payoff from emitting signal $\sigma_i$ is
\begin{equation}
\begin{split}
\pi_{\rho,i}(t+1)=\sum_{j=1}^{n}\rho_j(t)\Big[&x_i(t+1)x_j(t+1)R+x_j(t+1)\big(1-x_i(t+1)\big)S \\
&+\big(1-x_j(t+1)\big)x_i(t+1)T+\big(1-x_j(t+1)\big)\big(1-x_i(t+1)\big)P\Big]-c(\sigma_i).
\end{split}
\label{eq:mf_rho_base}
\end{equation}
Under the present opponent-conditioned model, when a focal individual emits $\sigma_i$ and encounters an opponent emitting $\sigma_j$, the focal responds to the opponent's signal and therefore cooperates with probability $x_j(t+1)$, while the opponent cooperates with probability $x_i(t+1)$. Defining $f_{\rho,i}(t+1)=e^{\beta\pi_{\rho,i}(t+1)}$, the signal frequencies evolve by selection,
\[
\rho_i^{\mathrm{sel}}(t+1)=\frac{\rho_i(t)f_{\rho,i}(t+1)}{\sum_{k=1}^{n}\rho_k(t)f_{\rho,k}(t+1)},
\]
followed by mutation,
\[
\rho_i^{\mathrm{mut}}(t+1)=\big(1-\mu_{\sigma}\big)\rho_i^{\mathrm{sel}}(t+1)+\frac{\mu_{\sigma}}{n},
\qquad
\mu_{\sigma}=\nu_{\sigma}\frac{d}{1+d},
\]
where $\mu_{\sigma}$ is the average drift toward the uniform distribution induced by the signal-mutation rule. Population stochasticity is then added as
\begin{equation}
\rho_i(t+1)=\rho_i^{\mathrm{mut}}(t+1)+\sigma_{\rho}\sqrt{\frac{\rho_i^{\mathrm{mut}}(t+1)\big(1-\rho_i^{\mathrm{mut}}(t+1)\big)}{N}}\,\xi_{\rho,i}(t),
\end{equation}
where $\xi_{\rho,i}(t)$ is again standard normal and $\sigma_{\rho}$ measures demographic noise. In the numerical implementation, negative entries are set to zero after this step and the resulting vector is renormalized so that $\sum_i \rho_i(t+1)=1$.

\begin{figure}[p]
\centering
\includegraphics[width=0.96\textwidth]{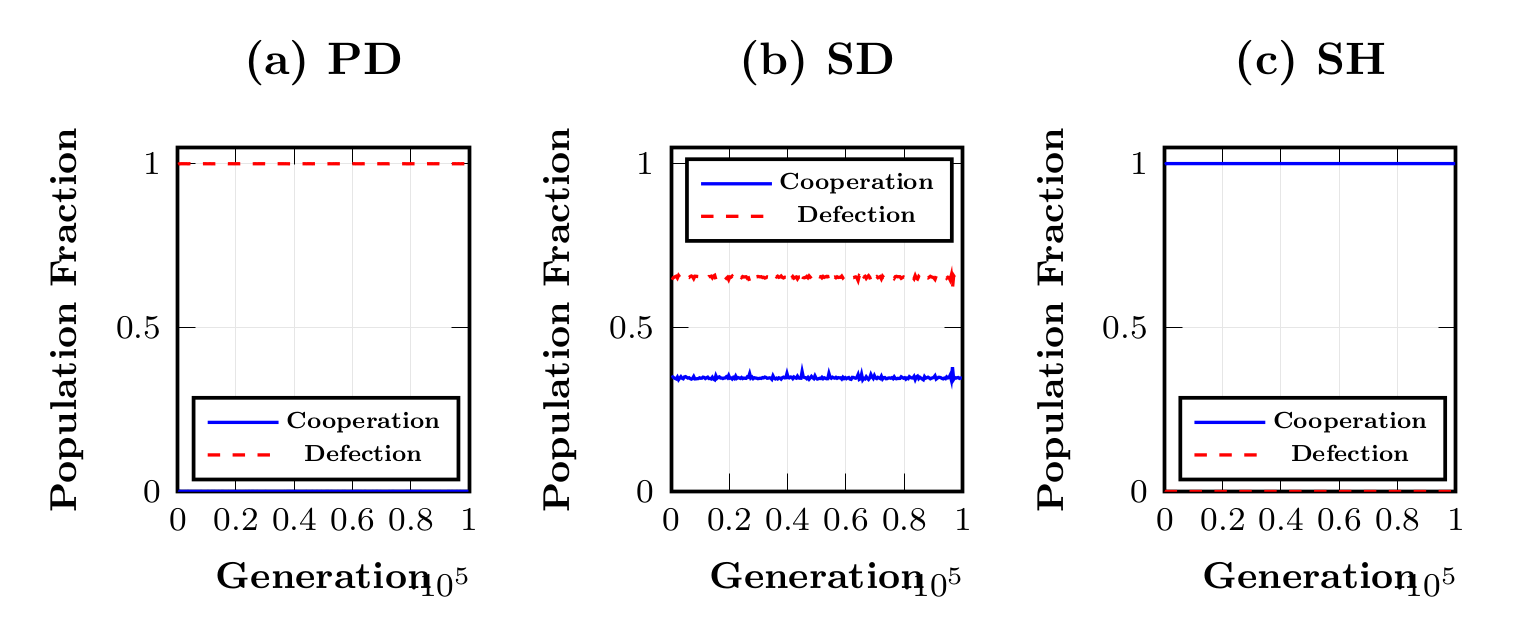}
  \caption{Stochastic numerical solution of the independence closure across the three canonical games. All panels use the default well-mixed simulation parameters from Table~\ref{tab:params}: $n=100$, $N=1800$, $\beta=1$, $\nu_s=\nu_{\sigma}=10^{-3}$, $n_s=10$, $d=0.2$, $c_{\max}=0.5$, $10^5$ generations, $\sigma_x=5$, and $\sigma_{\rho}=10$. In this implementation, the opponent-conditioned signal payoff uses Eq.~\eqref{eq:mf_rho_base}, strategy mutation acts on the post-selection state as $x\mapsto (1-\mu_s)x+\mu_s/2$ with $\mu_s=\nu_s n_s/n$, and signal mutation acts on the post-selection signal frequencies as $\rho\mapsto (1-\mu_\sigma)\rho+\mu_\sigma/n$ with $\mu_\sigma=\nu_{\sigma}d/(1+d)$. The stochastic independence closure still fails in PD, supports partial cooperation in SD, and yields near-complete cooperation in SH. The deterministic limit is shown in Supplementary Information, Fig.~S7. Parameters: $(R,S,T,P)=(3,0,5,1)$ in PD, $(3,1,5,0)$ in SD, and $(5,0,3,1)$ in SH.}
	\label{fig:independence_closure_stochastic_results}
\end{figure}

We solve the independence closure with the same finite-population stochastic terms used in the reduced model, using the default well-mixed simulation parameters from Table~\ref{tab:params}. The resulting dynamics are shown in Fig.~\ref{fig:independence_closure_stochastic_results}, and the associated signal statistics are shown in Fig.~\ref{fig:independence_closure_stochastic_signals}. In the Supplementary Information, Section~S5, we also show the deterministic limit (deterministic discrete replicator dynamics) obtained by setting $\sigma_x=\sigma_{\rho}=0$. The independence closure already reproduces an important asymmetry across games. In the PD, cooperation declines toward zero. This can be seen directly from the payoff difference
\[
\Delta\pi(t)=\pi_C(t)-\pi_D(t)=(R-T)\langle x\rangle_{\rho}+(S-P)\big(1-\langle x\rangle_{\rho}\big),
\qquad
\langle x\rangle_{\rho}=\sum_{j=1}^{n}\rho_j x_j.
\]
Because $T>R$ and $P>S$ in the PD, $\Delta\pi(t)$ is negative for every $\langle x\rangle_{\rho}\in[0,1]$, so the deterministic part of the dynamics drives all response classes toward defection. In Supplementary Information, Figs.~S7 and S8, we show that removing population stochasticity does not change this conclusion. The deterministic limit still collapses to defection in PD.

The SD and SH behave differently. In the SD, the sign of $\Delta\pi(t)$ depends on the current average level of cooperation, and the system settles into a partially cooperative state, consistent with Fig.~\ref{fig:independence_closure_stochastic_results}(b). In the SH, once the population moves above the coordination threshold, cooperation becomes self-reinforcing and the system approaches near-complete cooperation, as shown in Fig.~\ref{fig:independence_closure_stochastic_results}(c). The signal statistics in Fig.~\ref{fig:independence_closure_stochastic_signals} are consistent with this picture: signal usage is negatively related to cost and positively related to net signal payoff, with the strongest concentration in SH where the cooperative convention becomes most stable. Supplementary Information, Section~S5, shows that the deterministic limit preserves the same SD--SH contrast, but in the baseline closure it also drives signal usage much more strongly toward the cheapest signal. Population stochasticity therefore improves the empirical appearance of the signal-frequency distributions by keeping more signals in circulation, even though it does not alter the closure's qualitative success in SD and SH or its failure in PD. The main limitation of the independence closure is therefore specific and informative. Average population feedback is already sufficient to account qualitatively for SD and SH, but not for the partially cooperative PD regime. That failure points to the importance of correlations that are absent from this baseline description.

\subsubsection{Effective Protection of Rare Signals}\label{sec:rare_signal_protection_closure}
The independence closure misses a key feature of the PD dynamics. In the simulations, the cooperative value attached to a signal is often highest shortly after that signal appears, before defectors have become concentrated around it. The independence closure averages over signal classes at the population level and therefore smooths away this transient shelter. To represent that effect while retaining the same coarse-grained variables and the same generational structure, we extend the discrete-time stochastic replicator system by allowing rarity to confer a temporary advantage.

At generation $t$, the modified payoffs for responding to signal $\sigma_i$ are
\begin{align}
\pi_{C,i}'(t) &= \sum_{j=1}^{n}\rho_j(t)\big[x_j(t)R+\big(1-x_j(t)\big)S\big]+\eta\big(1-\rho_i(t)\big), \\
\pi_{D,i}'(t) &= \sum_{j=1}^{n}\rho_j(t)\big[x_j(t)T+\big(1-x_j(t)\big)P\big],
\end{align}
where $\eta>0$ measures the additional protection of cooperating against a rare signal. The linear form $\eta(1-\rho_i)$ is chosen as the simplest frequency-dependent correction that is strongest when a signal is rare and vanishes as it becomes common. With $f_{C,i}'(t)=e^{\beta\pi_{C,i}'(t)}$ and $f_{D,i}'(t)=e^{\beta\pi_{D,i}'(t)}$, the response profile again evolves first by selection,
\[
x_i^{\mathrm{sel}}(t+1)=\frac{x_i(t)f_{C,i}'(t)}{x_i(t)f_{C,i}'(t)+\big(1-x_i(t)\big)f_{D,i}'(t)},
\]
followed by mutation,
\[
x_i^{\mathrm{mut}}(t+1)=\big(1-\mu_s\big)x_i^{\mathrm{sel}}(t+1)+\frac{\mu_s}{2},
\qquad
\mu_s=\nu_s\frac{n_s}{n},
\]
and population stochasticity,
\begin{equation}
x_i(t+1)=x_i^{\mathrm{mut}}(t+1)+\sigma_x\sqrt{\frac{x_i^{\mathrm{mut}}(t+1)\big(1-x_i^{\mathrm{mut}}(t+1)\big)}{N}}\,\xi_{x,i}(t).
\label{eq:mf_x_corr}
\end{equation}
As in the independence closure, $\xi_{x,i}(t)$ is a standard normal draw and the numerical implementation truncates values back to $[0,1]$ after the stochastic step. When $\sigma_x=0$ and $\sigma_\rho=0$, the deterministic part of the model reduces to a discrete-time replicator-mutator system with rarity-modified payoffs.

The same idea is applied to signal production. Under the opponent-conditioned model, the payoff from emitting signal $\sigma_i$ is
\begin{equation}
\begin{split}
\pi_{\rho,i}'(t+1)=\sum_{j=1}^{n}\rho_j(t)\Big[&x_i(t+1)x_j(t+1)R+x_j(t+1)\big(1-x_i(t+1)\big)S \\
&+\big(1-x_j(t+1)\big)x_i(t+1)T+\big(1-x_j(t+1)\big)\big(1-x_i(t+1)\big)P\Big] \\
&-c(\sigma_i)+\delta\big(1-\rho_i(t)\big),
\end{split}
\label{eq:mf_rho_corr}
\end{equation}
where $\delta>0$ is the signal-side rarity bonus. Defining $f_{\rho,i}'(t+1)=e^{\beta\pi_{\rho,i}'(t+1)}$, the signal frequencies first evolve by selection,
\[
\rho_i^{\mathrm{sel}}(t+1)=\frac{\rho_i(t)f_{\rho,i}'(t+1)}{\sum_{k=1}^{n}\rho_k(t)f_{\rho,k}'(t+1)},
\]
then by mutation,
\[
\rho_i^{\mathrm{mut}}(t+1)=\big(1-\mu_{\sigma}\big)\rho_i^{\mathrm{sel}}(t+1)+\frac{\mu_{\sigma}}{n},
\qquad
\mu_{\sigma}=\nu_{\sigma}\frac{d}{1+d},
\]
and finally by population stochasticity,
\begin{equation}
\rho_i(t+1)=\rho_i^{\mathrm{mut}}(t+1)+\sigma_{\rho}\sqrt{\frac{\rho_i^{\mathrm{mut}}(t+1)\big(1-\rho_i^{\mathrm{mut}}(t+1)\big)}{N}}\,\xi_{\rho,i}(t).
\end{equation}
Here $\xi_{\rho,i}(t)$ is again standard normal. In the numerical implementation, negative entries are set to zero after the stochastic step and the resulting vector is renormalized so that $\sum_i\rho_i(t+1)=1$.

This second closure should be interpreted as a reduced representation of the missing frequency dependence rather than as a full derivation of the agent-based process. Its role is to test whether temporary protection of uncommon signals is sufficient to recover the partially cooperative PD regime that the independence closure misses.

\begin{figure}[p]
\centering
\includegraphics[width=0.96\textwidth]{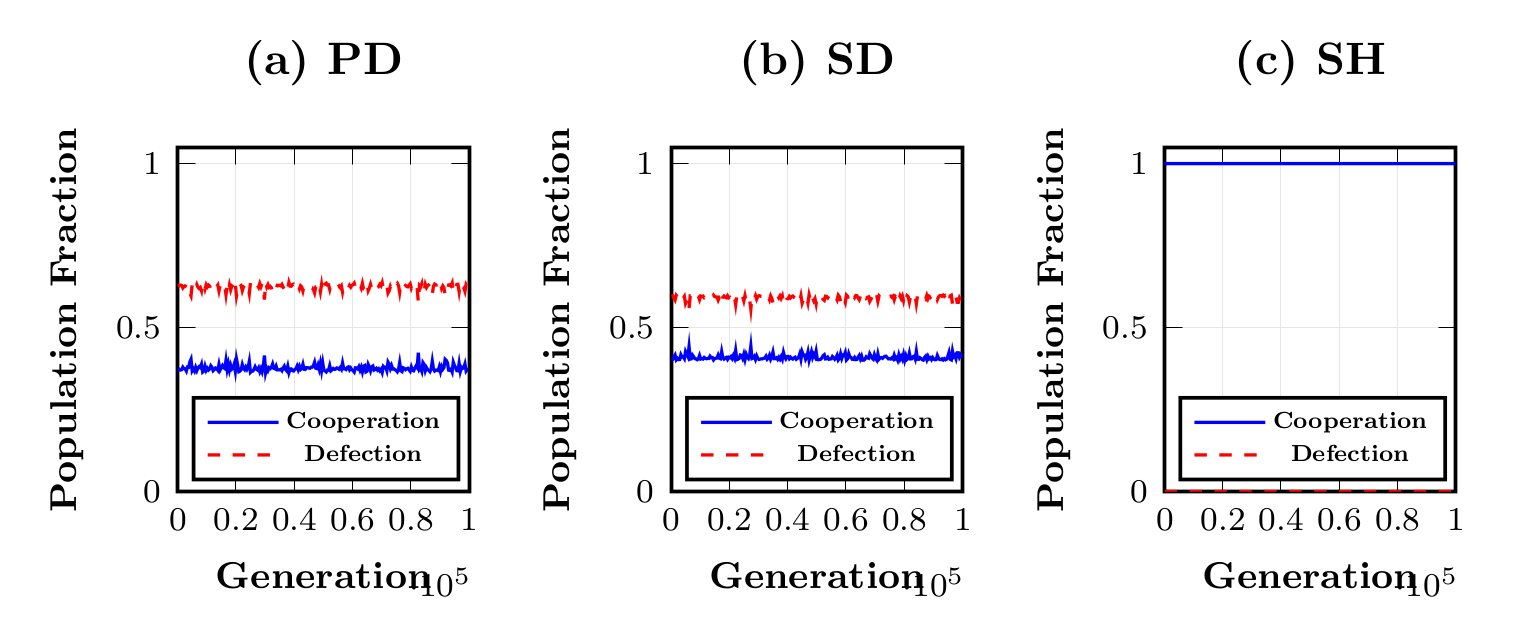}
  \caption{Stochastic numerical solution of the rare-signal-protection closure across the three canonical games. All panels use the default well-mixed simulation parameters from Table~\ref{tab:params}: $n=100$, $N=1800$, $\beta=1$, $\nu_s=\nu_{\sigma}=10^{-3}$, $n_s=10$, $d=0.2$, $c_{\max}=0.5$, $10^5$ generations, $\sigma_x=5$, and $\sigma_{\rho}=10$. The rarity parameters are tuned separately by game while keeping all non-rarity parameters fixed: for PD, $\eta=1.4$ and $\delta=0.3$; for SD, $\eta=0.2$ and $\delta=0.05$; for SH, $\eta=0.1$ and $\delta=0.05$. As in Fig.~\ref{fig:independence_closure_stochastic_results}, the implementation uses the opponent-conditioned signal payoff and post-selection mutation in both the strategy and signal updates. The rare-signal-protection closure preserves partial cooperation in PD and SD, approaches near-complete cooperation in SH, and yields signal statistics that are substantially closer to the simulation results than the baseline closure. Parameters: $(R,S,T,P)=(3,0,5,1)$ in PD, $(3,1,5,0)$ in SD, and $(5,0,3,1)$ in SH.}
	\label{fig:rare_signal_protection_stochastic_results}
\end{figure}

\begin{figure}[p]
\centering
\includegraphics[width=0.96\textwidth]{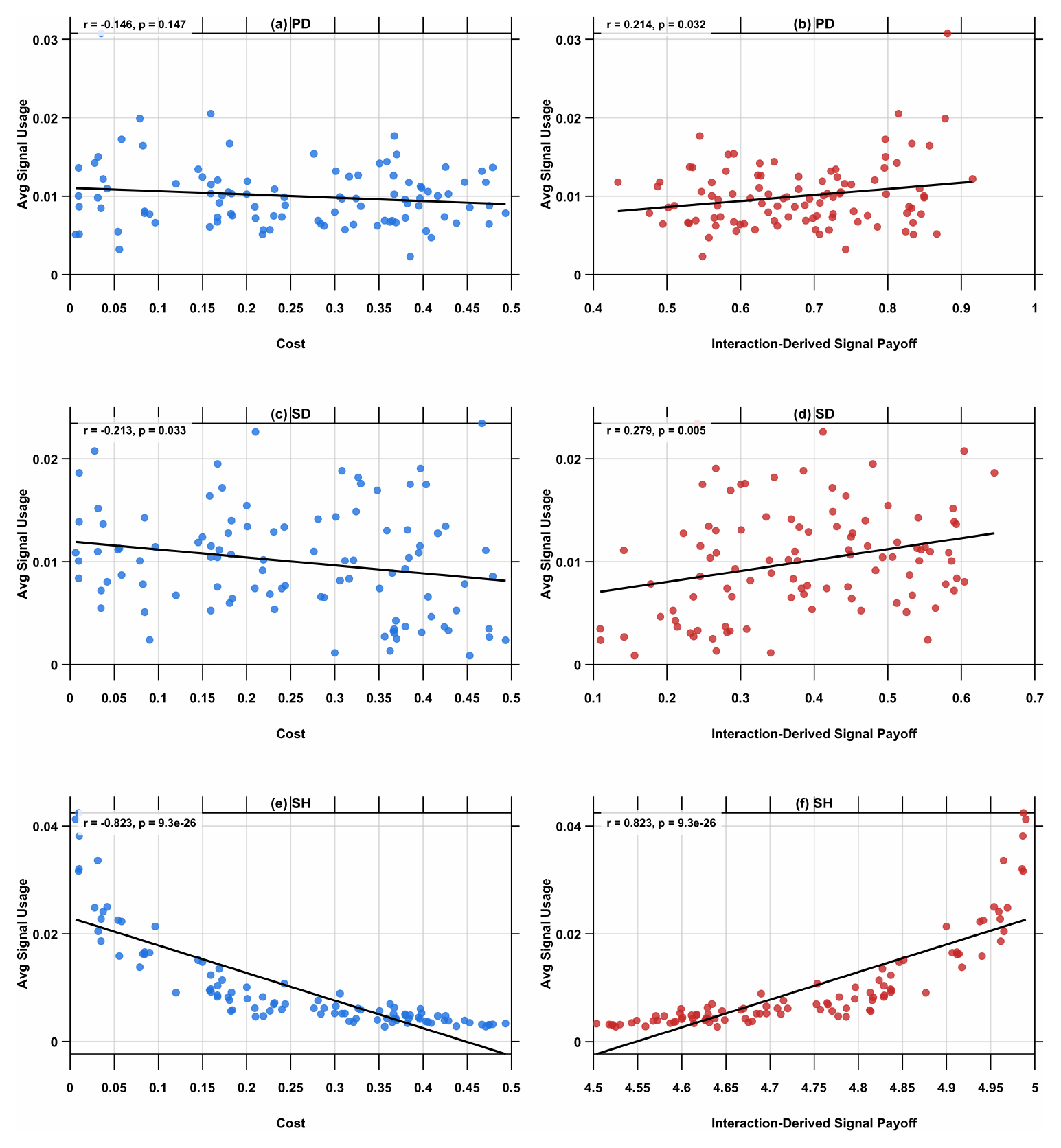}
  \caption{Signal statistics for the stochastic numerical solution of the rare-signal-protection closure. The non-rarity parameters are the same as in Fig.~\ref{fig:rare_signal_protection_stochastic_results}. The rarity parameters are $\eta=1.4$, $\delta=0.3$ in PD, $\eta=0.2$, $\delta=0.05$ in SD, and $\eta=0.1$, $\delta=0.05$ in SH. Each point corresponds to one signal, and the averages are taken over the last $2\times 10^4$ generations. Panels (a,c,e) plot average signal usage against cost. Panels (b,d,f) plot average signal usage against the interaction-derived signal payoff, namely the mean signal payoff excluding the explicit rarity bonus, so that the horizontal axis is directly comparable to the simulation panels. The black line in each panel is the least-squares trend, and the inset gives the Pearson correlation coefficient and its two-sided $p$ value. Parameters: $n=100$, $N=1800$, $\beta=1$, $\nu_s=\nu_{\sigma}=10^{-3}$, $n_s=10$, $d=0.2$, $c_{\max}=0.5$, $10^5$ generations, $\sigma_x=5$, $\sigma_{\rho}=10$, with $(\eta,\delta)=(1.4,0.3)$ in PD, $(0.2,0.05)$ in SD, and $(0.1,0.05)$ in SH, and $(R,S,T,P)=(3,0,5,1)$ in PD, $(3,1,5,0)$ in SD, and $(5,0,3,1)$ in SH.}
	\label{fig:rare_signal_protection_stochastic_signals}
\end{figure}

We solve the rare-signal-protection closure with the same finite-population stochastic terms used in the reduced model, again using the default well-mixed simulation parameters from Table~\ref{tab:params}. The resulting dynamics are shown in Fig.~\ref{fig:rare_signal_protection_stochastic_results}, and the associated signal statistics are shown in Fig.~\ref{fig:rare_signal_protection_stochastic_signals}. The non-rarity parameters are held fixed across games, while the rarity parameters are allowed to vary modestly because the omitted correlations summarized by the reduced correction are not equally strong in PD, SD, and SH. In the numerical solutions shown here, $(\eta,\delta)=(1.4,0.3)$ for PD, $(0.2,0.05)$ for SD, and $(0.1,0.05)$ for SH. In the Supplementary Information, Section~S5, we also show the deterministic limit obtained by setting $\sigma_x=\sigma_{\rho}=0$.

The numerical results recover the main asymmetry across games. In the PD, cooperation no longer collapses to zero. Instead, the rarity terms allow newly emerged signals to sustain temporary cooperative niches before exploitation catches up, and that repeated creation of short-lived protected niches is enough to maintain a partially cooperative regime. In the SD, only a weak rarity correction is required because the underlying payoff structure already permits coexistence between cooperation and defection. The rarity terms therefore modify the quantitative balance more than the qualitative state. In the SH, the cooperative outcome is already favored once coordination becomes established, so again only a weak correction is needed; the main effect is to accelerate concentration around the prevailing cooperative convention. The point of the correction is therefore not exact time-series fitting, but identification of the missing ingredient: temporary protection of uncommon signals is sufficient to recover the qualitative PD behavior that the independence closure misses. Supplementary Information, Figs.~S9 and S10, show that this conclusion survives even without population stochasticity.

The signal statistics in Fig.~\ref{fig:rare_signal_protection_stochastic_signals} are consistent with this interpretation. Across all three games, signal usage decreases with cost and increases with interaction-derived signal payoff, with the sharpest concentration again occurring in SH. This stronger concentration in SH reflects the fact that once a cooperative convention is established, there is less continuing conflict over the response attached to a signal. In PD and SD, by contrast, exploitation continues to erode successful conventions, so signal usage remains more dispersed across signal space. For the payoff panels, we plot the interaction-derived signal payoff excluding the explicit rarity term, so that the horizontal axis remains directly comparable to the simulation figures. Supplementary Information, Section~S5, shows that the deterministic rare-signal-protection closure already produces a much more informative cost and payoff trend than the deterministic independence closure. Population stochasticity still broadens the clouds, but here it is not doing the main conceptual work. The stronger signal pattern already follows from the frequency-dependent protection of uncommon signals itself. Taken together, the results show that a minimal frequency-dependent protection of rare signals is sufficient to recover the missing PD behavior while preserving the qualitative SD and SH patterns already captured by the simpler closure.

		\section{Discussion and Conclusion}

We showed that signal-response feedback can support cooperation across a broad range of strategic settings, but it does so in different ways in different games. In well-mixed populations, the mechanism sustains partial cooperation in PD and SD and drives near-complete cooperation in SH. On the lattice, cooperation rises further because local clustering and signal-based discrimination reinforce one another. In fluctuating strategic environments, the same mechanism adapts smoothly: PD-SD mixtures remain close to the partially cooperative regime, while PD-SH mixtures interpolate between near-full cooperation and the harsher PD regime.

The signal statistics clarify why raw cost alone is a poor predictor of prevalence. Across games, low-cost signals are often favored, but what really determines persistence is whether a signal continues to attract cooperation. Once a signal becomes too common and begins to attract defection, its effective fitness falls and its prevalence declines. The results are therefore better described as competition among signals for favorable response conventions than as a simple ranking by production cost.

The mean-field analysis helps separate what can be explained by the baseline stochastic model from what cannot. The decoupled closure already accounts for the success of the mechanism in SD and SH, but it fails in the PD because one-shot defection remains the dominant response under purely average mixing. The rare-signal-protection closure shows that adding a minimal rarity-dependent protection for uncommon signals is already enough to recover partial cooperation in the PD while preserving the SD and SH patterns. The reduced theory should therefore be read as a mechanism-isolating approximation: it is informative because it identifies which part of the simulation dynamics must be retained once average mixing is no longer enough.

Relative to \cite{salahshour2019}, the present results clarify the scope conditions of the mechanism. The earlier work identified partial cooperation in a well-mixed setting. Here the same line of inquiry is followed across multiple games and population structures, and in a simplified behavioral formulation where actions depend only on the opponent's signal. Taken together, the simulations and reduced theory show where the mechanism carries over, where it becomes stronger, and why its PD version depends on correlations that an average-mixing description necessarily misses.

While our study demonstrates that the coevolution of costly signals and cooperative strategies is a plausible route to the evolution of both costly signals and costly cooperative strategies, the implications of our study depend on the empirical class of costly signals being considered. Handicap-based accounts \cite{zahavi1975,grafen1990}, honest-signaling accounts \cite{gintis2001,fehrler2013,raihani2015}, and competitive-altruism accounts \cite{roberts1998,barclaywiller2007} remain especially plausible when displays track relatively stable differences in quality, commitment, or trustworthiness, and when receivers use those displays to choose long-term partners, mates, or allies. In those settings, the main role of cost is to help maintain reliability or to make generosity informative.

The mechanism studied here is more plausible in a different empirical regime: one in which cues are socially learned, partly arbitrary, or only weakly tied to intrinsic quality, but still become behaviorally important because they organize expectations about whom to help, trust, or avoid. Work on ethnic markers, symbolic labels, group identity, and language-like coordination shows that such cues can reshape cooperation because they sort interactions and support shared response conventions \cite{riolo2001,jansen2006,mcelreath2003,efferson2008,chenli2009,salahshour2020language}. In those settings, costly signaling need not be understood primarily as honest revelation. It can instead be understood as a way of stabilizing a convention long enough for cooperation to become locally advantageous. In that sense, the present model is most relevant for cases in which an observable badge, style, ritualized display, or linguistic cue acquires value through the cooperation it elicits, rather than through a fixed one-to-one link with hidden quality \cite{salahshour2019}.

This perspective also suggests different empirical expectations. Relative to handicap-style accounts, it predicts weaker links between signal cost and underlying quality, stronger dependence on local convention, and more rapid turnover once an initially successful cue is imitated widely enough to attract exploitation. Relative to explanations based only on kinship or static clustering, it predicts that visible markers can matter in their own right by defining who is treated as a preferred partner within the same population structure. These predictions are broadly consistent with the literatures on symbolic markers and convention-based assortment, though testing them directly would require longitudinal data on how cooperative cues spread, concentrate locally, and lose value over time \cite{mcelreath2003,efferson2008,chenli2009,salahshour2019}.

Seen this way, the present account complements rather than replaces alternative signaling theories. Honest-quality and partner-choice models remain the natural benchmark when signal reliability is anchored in stable differences among individuals \cite{gintis2001,roberts1998,barclaywiller2007,fu2008partnerchoice}. The endogenous-convention account developed here is more appropriate when the main empirical problem is not truthful revelation of type, but the emergence and erosion of locally meaningful markers that transiently support cooperative niches. That distinction may help explain why some costly displays appear stable and informative, whereas others are context dependent, strategically imitated, and vulnerable to turnover \cite{lachmann2001,szamado2011,szamado2023,mcelreath2003,efferson2008}.

The model is deliberately stylized: signals are arbitrary, interactions are one-shot, and receivers condition behavior only on observed cues. Its value is therefore not that it captures every empirical form of costly signaling, but that it isolates a mechanism by which signaling cost can become worthwhile through the cooperative responses it organizes. Future work could extend the same framework in several directions. More realistic networks would show whether the effect depends mainly on local clustering or on finer community structure. A richer signal space could test whether the same turnover persists when signals are continuous or combinatorial. It would also be useful to derive a more explicit theory of the genotype-level correlations that the present mean-field correction summarizes only phenomenologically.

\section*{Acknowledgements}

M.S. acknowledges funding from German Research Foundation (DFG---Deutsche Forschungsgemeinschaft) under Germany's Excellence Strategy---EXC 2117--422037984.

\bibliographystyle{plain}

\begin{figure}[p]
\centering
\includegraphics[width=0.96\textwidth]{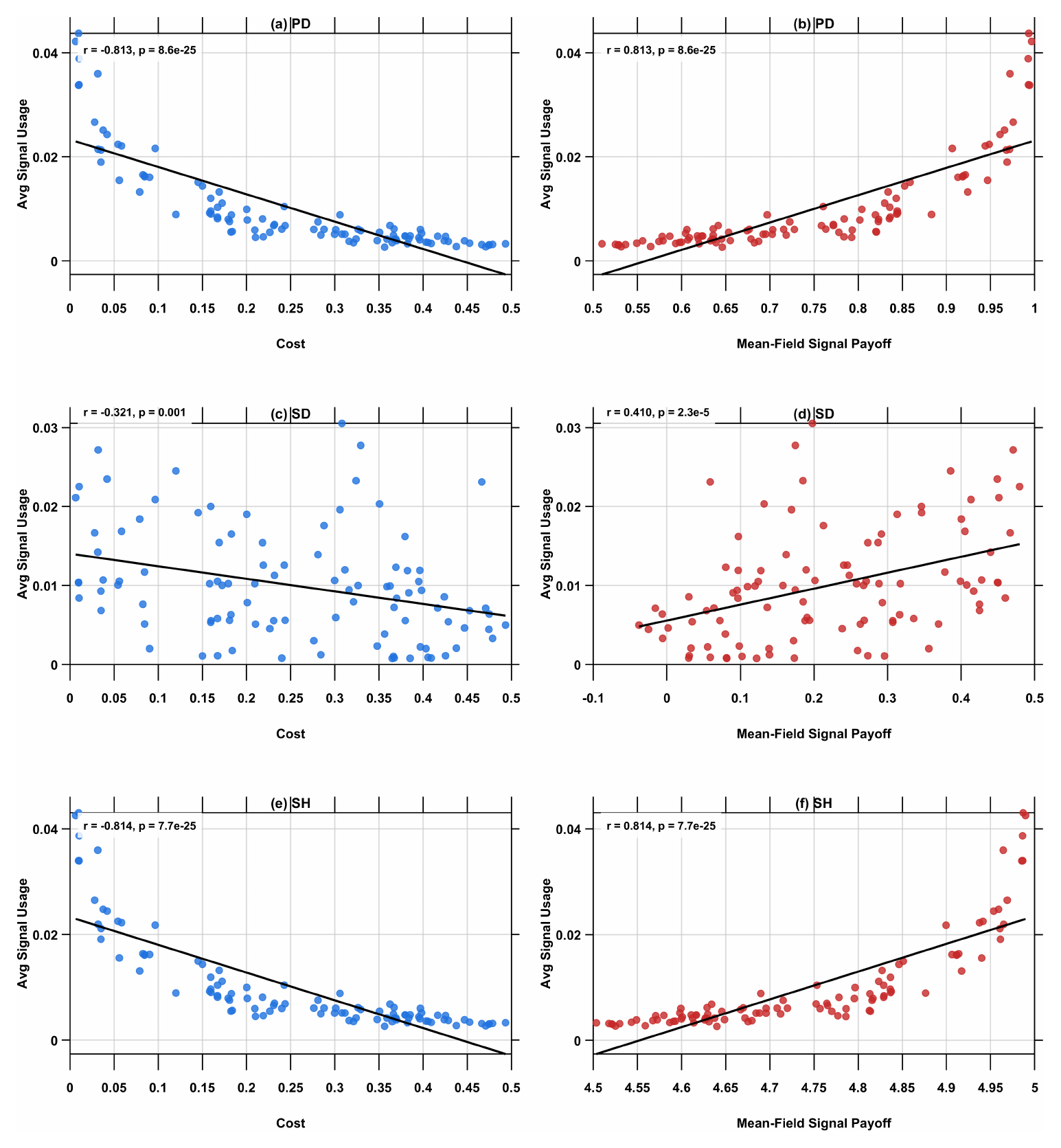}
  \caption{Signal statistics for the stochastic numerical solution of the independence closure. The parameters are the same as in Fig.~\ref{fig:independence_closure_stochastic_results}. Each point corresponds to one signal, and the averages are taken over the last $2\times 10^4$ generations of the run. Panels (a,c,e) plot average signal usage against signal cost, while panels (b,d,f) plot average signal usage against the mean-field signal payoff, averaged over the same time window. The black line in each panel is the least-squares trend, and the inset gives the Pearson correlation coefficient and its two-sided $p$ value. Parameters: $n=100$, $N=1800$, $\beta=1$, $\nu_s=\nu_{\sigma}=10^{-3}$, $n_s=10$, $d=0.2$, $c_{\max}=0.5$, $10^5$ generations, $\sigma_x=5$, $\sigma_{\rho}=10$, and $(R,S,T,P)=(3,0,5,1)$ in PD, $(3,1,5,0)$ in SD, and $(5,0,3,1)$ in SH.}
	\label{fig:independence_closure_stochastic_signals}
\end{figure}

\clearpage

\end{document}